\documentclass[aip,jcp,a4paper,reprint,floatfix,superscriptaddress]{revtex4-1}

\usepackage[caption=false]{subfig}
\usepackage{graphicx}
\usepackage{color}
\usepackage{amsmath}
\usepackage{dsfont}
\usepackage{xspace}
\usepackage{bm}
\usepackage{txfonts}
\usepackage{xr}

\definecolor{delftdark}{rgb}{.6, .2, .2}

\usepackage[plainpages=false, pdfpagelabels,
    pdfpagemode=UseNone,
    pdfstartview=FitH,
    colorlinks=true,
    linkcolor=delftdark,
    citecolor=delftdark,
    a4paper=true
    pdfstartpage=1,
    pdfauthor={Verzijl, Celis Gil, Perrin, Dulic, van der Zant and Thijssen},
    pdfsubject={Image Effects in Transport at Metal-Molecule Interfaces},
    pdftitle={},
]{hyperref}

\newcommand{\rijI}{|{\bm r}_i-{\bm r}_{j}^I|}

\newcommand{\etal}{\emph{et al.}\xspace}
\newcommand{\eg}{\emph{e.g.}\xspace}

\newcommand{\bfr}{{\bm r}}

\begin{document}

\title{Image Effects in Transport at Metal-Molecule Interfaces}

\author{C.J.O. Verzijl}
\affiliation{Kavli Institute of Nanoscience, Delft University of Technology, 2628 CJ Delft, The Netherlands}
\author{J.A. Celis Gil}
\affiliation{Kavli Institute of Nanoscience, Delft University of Technology, 2628 CJ Delft, The Netherlands}
\email{J.A.CelisGil@TUDelft.nl}
\author{M.L. Perrin}
\affiliation{Kavli Institute of Nanoscience, Delft University of Technology, 2628 CJ Delft, The Netherlands}
\author{D. Duli\'c}
\affiliation{Departamento de F\'isica, Facultad de Ciencias F\'isicas y Matem\'aticas, Universidad de Chile, Santiago de Chile, Chile.}
\author{H.S.J. van der Zant}
\affiliation{Kavli Institute of Nanoscience, Delft University of Technology, 2628 CJ Delft, The Netherlands}
\author{J.M. Thijssen}
\affiliation{Kavli Institute of Nanoscience, Delft University of Technology, 2628 CJ Delft, The Netherlands}

\date{\today}


\begin{abstract}
We present a method for incorporating image-charge effects into the description of charge transport through molecular devices. A simple model allows us to calculate the adjustment of the transport levels, due to the polarization of the electrodes as charge is added to and removed from the molecule. For this, we use the charge distributions of the molecule between two metal electrodes in several charge states, rather than in gas phase, as obtained from a DFT-based transport code.
This enables us to efficiently model level shifts and gap renormalization caused by image-charge effects, which are essential for understanding molecular transport experiments. We apply the method to benzene di-amine molecules and compare our results with the standard approach based on gas phase charges. Finally, we give a detailed account of the application of our approach to porphyrin-derivative devices recently studied experimentally by Perrin \etal, which demonstrates the importance of accounting for image-charge effects when modeling transport through molecular junctions.
\end{abstract}

\maketitle

\section{Introduction}

Understanding the physics determining charge transport at interfaces between metal electrodes and molecules is key to the advancement of the field of molecular electronics. In a molecular device, the alignment of frontier molecular orbital levels relative to the metals' Fermi energies determines the contribution of the different channels available for transport. Due to their proximity to the electrodes, the levels themselves are shifted relative to those of the molecule in gas phase, and may hybridize with electrode levels as well. Together, level alignment and hybridization determine electron transport in the molecular junction.

In this paper we describe an approach to investigating these effects based on density-functional theory (DFT)\cite{Jones1989} and the non-equilibrium Green's functions (NEGF) formalism.\cite{Meir1992,Datta2000,Brandbyge2002,Stokbro2003a,Evers2003,Rocha2006,Rothig2006,Arnold2007,Verzijl2012} 
DFT is frequently used in calculations of charge transport because of its efficiency, and because computationally it scales well to realistic nanoscale junction sizes. It does suffer from a few drawbacks, however, the most important of which are poor predictions of one- and two-particle excitations.\cite{Jones1989,Burke2012}  
The reason for the failure of DFT to predict excitation energies from a single neutral-state calculation is mainly due to the inclusion of spurious self-interactions,\cite{Perdew1981,Toher2005} and the omission of dynamic polarization effects.\cite{Hybertsen1986,Neaton2006}
Both effects are captured in GW calculations,\cite{Aryasetiawan1998,Neaton2006,Thygesen2009} usually within the COHSEX approach,\cite{Hybertsen1986} and time-dependent density-functional theory (TDDFT).\cite{Stefanucci2004,Kurth2005,Perfetto2010} However, these are computationally expensive and not (yet) feasible except for very small molecules, in contrast to DFT-based approaches.

Approximate methods have been proposed and used with some success to address the shortcomings of DFT in predicting excitations. These include the use of a scissors-operator\cite{Quek2007,Mowbray2008} and simple image-charge models based on atomic charges,\cite{Hedegaard2005,Neaton2006,Kaasbjerg2008,Mowbray2008}
used to address the location of resonant levels in the transport region of the molecular device. 

In this paper, we focus on the latter and argue that image charges used in an electrostatic-energy calculation should be taken from the molecule in the presence of contacts rather than from the gas phase.
In section \ref{models} we provide a brief introduction to interface effects, and outline our method for the calculation of the image-charge effects.
In section \ref{Sec:BDA}, we apply our method to the 1,4-benzenediamine molecule between two gold electrodes and compare it to other approaches which have appeared in the literature\cite{Quek2007,Kaasbjerg2008,Mowbray2008}. Then, in section~\ref{Sec:ZnTPP}, we cover the application of our method to Zn-porphyrin devices studied in recent experiments by Perrin \etal \cite{Perrin2013}, in which image-charge effects play an important role.

\section{Theoretical model}\label{models}

\begin{figure*}
\subfloat[]{
   \includegraphics[width=1.2\columnwidth]{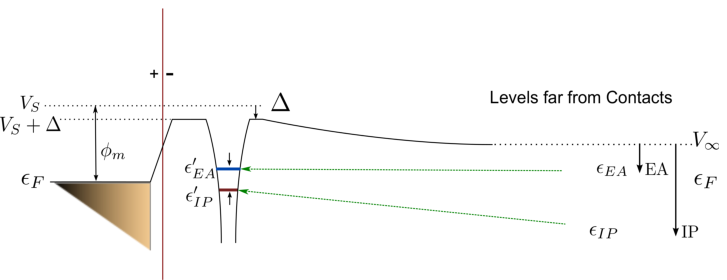}\label{fg:shiftstoon0}
   }\\
\subfloat[]{
   \includegraphics[width=1.0\columnwidth]{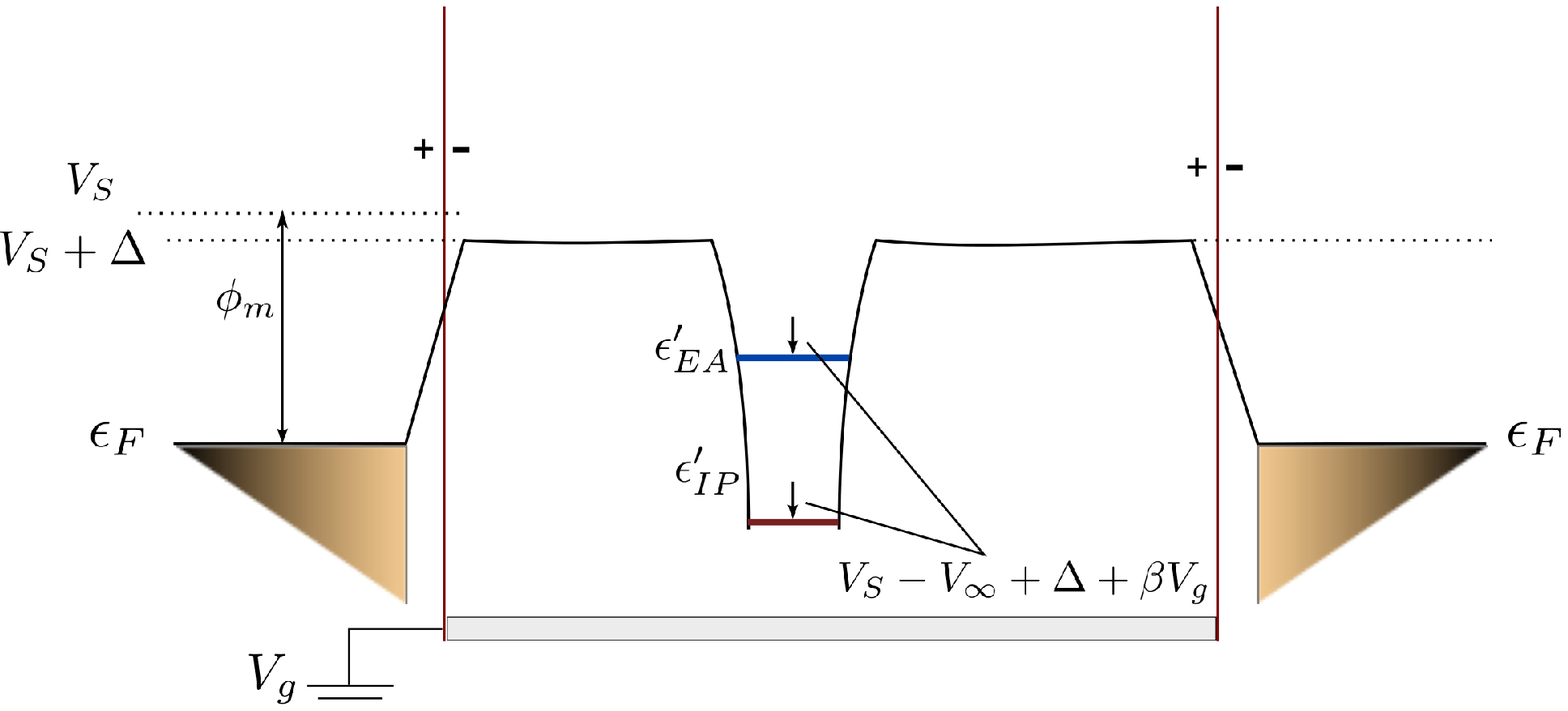}\label{fg:shiftstoon1} 
   }
\subfloat[]{
   \raisebox{.4cm}{\includegraphics[width=1.0\columnwidth]{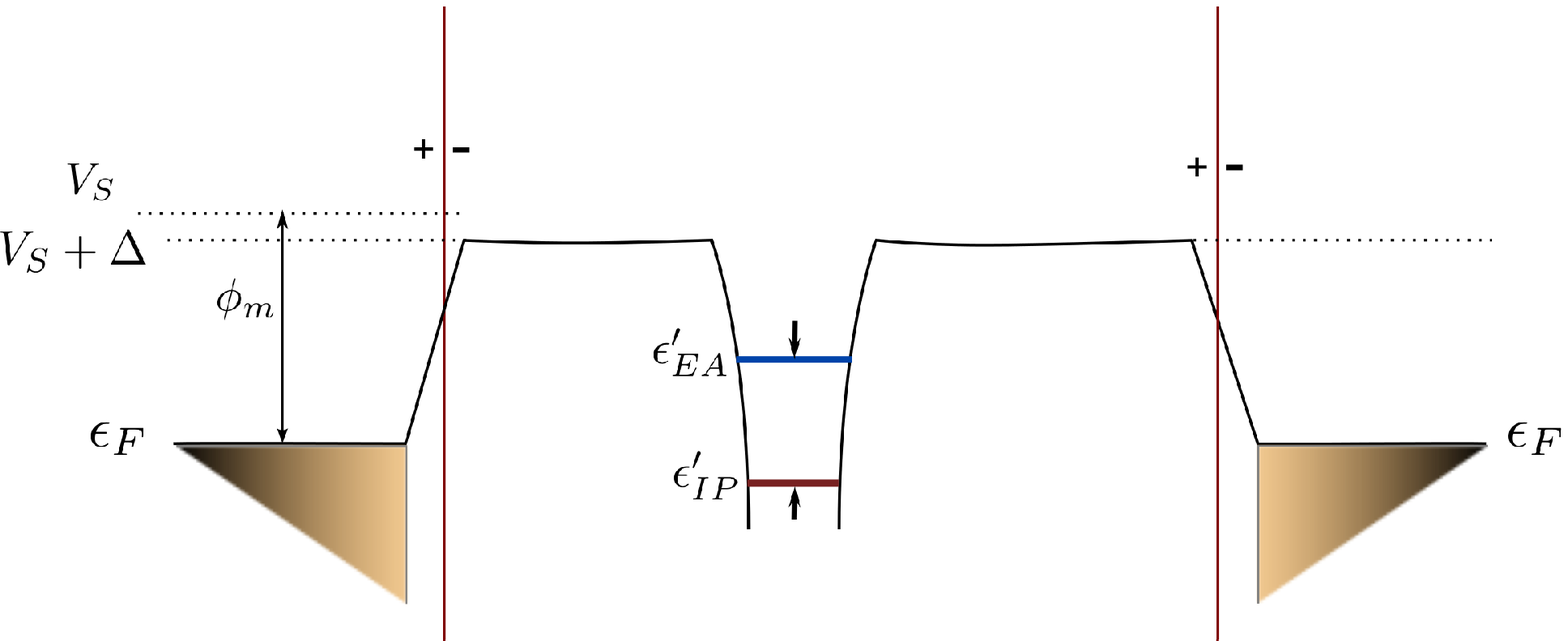}\label{fg:shiftstoon2}} 
   }
\caption{\textbf{Energy landscape during the formation of a metal-molecule interface.} 
(a) Combined rigid and dynamical image-charge effects on molecular levels at a single interface, relative to the molecule in isolation far away. These are a superposition of (b) and (c), where in (b) the surface dipole (shaded red/green) raises the background potential by $V_s-V_\infty$. The static image-charge effect, intrinsic molecular and interface dipoles shift the molecular levels back by $\Delta$, while electrostatic gating shifts by $\beta V_g$. (c) Levels are also subject to renormalization of the gap between the electron affinity $\epsilon_{EA}$ and the ionization potential $\epsilon_{IP}$ levels, where the prime indicates the position after the shift. 
}\label{fg:landscape}
\end{figure*}

We illustrate the most important physical effects as a molecule approaches a clean metal surface in Fig.~\ref{fg:landscape}, following Ishii \etal\cite{Ishii2000} 
The shift of the levels occurring when a molecule is moved towards a metal surface can be divided into two classes. The first type of shift is due to a change 
in the background potential induced by the proximity of the metal and, although different molecular orbitals may shift differently from their gas phase values, generally the \emph{direction} of these shifts is the same for all, and the differences between them are rather small.  Therefore, we denote these shifts as ``rigid''. Usually, these shifts are upward with respect to the gas phase.  

These background effects have their origin in the so-called ``push back'', or ``pillow'' effect, which refers to the reduction of the spill-out of electronic charge from the surface occurring for a clean metal surface. This spill-out results in a surface dipole which increases the work function.
As the push back effect reduces this spill-out (the molecule pushes the electronic charge back into the metal) it \emph{lowers} the work function.\cite{Seki1997,Ishii1999,Ishii2000,Vazquez2004a,Vazquez2004b}
A second mechanism resulting in a uniform shift of the orbital levels is charge transfer as a result of chemisorption, which also
changes the surface dipole. Finally, the charge distribution on the 
(possibly neutral) molecule generates an image charge distribution in the metal. The potential between the charges on the 
molecule and their images then results in a shift. The uniform shift resulting from all three mechanisms is denoted as $\Delta$ 
-- see Fig.~\ref{fg:shiftstoon0}.  
Oszwaldowski \etal have introduced a many-body method based on DFT\cite{Oszwaldowski2003} for capturing some of this dependence, deriving from dipole and pillow effects. 

The length scale over which the changes to the energy landscape due to a surface dipole layer take place is related to the lateral extent of the surface dipole layer formed at the metal surface. This is typically the scale of the electrode in a mechanically controlled break-junction (MCBJ) experiment, which is of the order of $5$ nm.\footnote{Estimated from the fits of the junction area in Perrin \etal's experiments\cite{Perrin2013}, which fit this as $28$ nm$^2$ and considered the range of $10-50$ nm$^2$ as representative.}
The magnitude of $\Delta$ is suggested by the measurements summarized by Ishii \etal: roughly $0.5-1$ eV, typically a negative correction on an Au substrate. 
Measurements by Koch \etal\cite{Duhm2008,Broker2010,Niederhausen2011} on thin-films with different molecules support these considerations: they find a constant-shift region very near the interface, followed by a linear shift of $\sim1$ eV over a range of roughly $8$\AA\xspace beyond which a regime with constant $\Delta$ sets in.

In addition to this rigid shift, there is a shift which is very different between occupied and unoccupied levels, causing the transport gap between them to close (``renormalize'') as the molecule approaches the surface. The upward shift of occupied levels is caused by the fact that
an electron moving away from the molecule leaves a positive hole behind. The electrostatic force needed to overcome when moving an electron from the molecule to 
infinity is for a substantial part responsible for the ionization potential of the molecule. If the molecule is close to a metal, removing electron 
from it will not only leave a positive hole behind, but also a negative image charge in the metal bulk. This \emph{reduces} the binding energy and 
therefore the ionization potential (IP).

Adding an electron to the molecule usually costs energy -- this is the \emph{addition energy}, AE. Close to a metal surface, however, the additional 
electron feels an attraction from the positive image charge it creates in the metal. Therefore, also the addition energy is reduced. We see that the
gap between the occupied and unoccupied levels therefore shrinks; this is denoted the gap renormalisation. In a transport junction, this gap 
is called the \emph{transport gap}. It should be noted that the above discussion relies on weak coupling between the molecule and the metal, implying
a preferentially integer electron occupation on the molecule. The rigid shift and the gap renormalization are effect is schematically represented in Fig.~\ref{fg:shiftstoon0}.

Gap renormalization has been studied extensively in the literature \cite{Quek2007,Neaton2006,Hybertsen1986,Hybertsen2008,Thygesen2009}. 
It was shown by Neaton \etal\cite{Neaton2006} that for small molecules this effect can be well fitted by an image-potential of the form 
$-\frac{1}{4z}$ (with $z$ the distance to the image plane). 
These effects are important in many nanoscale molecular systems, as has been argued on both experimental\cite{Kubatkin2003,Hedegaard2005,Bruot2012} and theoretical grounds \cite{Neaton2006,Quek2007,Mowbray2008,Kaasbjerg2008,Hybertsen2008,Thygesen2009} and are crucial for understanding and designing future molecular devices.

Electrostatic relaxation upon changing the charge on a molecule is not appropriately accounted for in DFT calculations, and in particular, it is missed in DFT-based NEGF calculations commonly used in studying single-molecule charge transport.
\cite{Brandbyge2002,Stokbro2003a,Rocha2006,Evers2006,Verzijl2012} We note that there are two types of relaxation: the first is the relaxation of the 
resident electrons on the molecule upon removing or adding an electron. These effects are responsible for the difference that is observed between
the HOMO (highest occupied level found in a DFT calculation) and the ionization potential, and similar for the AE and the LUMO.
This notion has led to applying the molecular shifts to the transport junction, one of the ingredients in the `scissors operator' approach.
\cite{Quek2007,Neaton2006,Thygesen2009} We have however seen that the image charges in the metal also shift the IP and AE, and these effects 
vary with the distance from the molecule to the metal contacts. This distance dependence is accessible in experiments (see section~\ref{experiments}) and the focus of this paper. 
Kaasbjerg and Flensberg\cite{Kaasbjerg2008} have addressed this effect and reported substantial gap reductions, and 
even dramatic ones in the presence of a gate. GW calculations in principle address such polarization effects, but these require very heavy computer resources
even for small systems.
Here, we use classical electrostatic calculations to address polarization effects due to the contacts, based on charge distributions obtained
from DFT calculations in different charge states. We note that DFT is designed for and has proven to be reliable for calculating ground state properties, and these
are the only ones used in our calculations.

\begin{figure}
\includegraphics[width=\columnwidth]{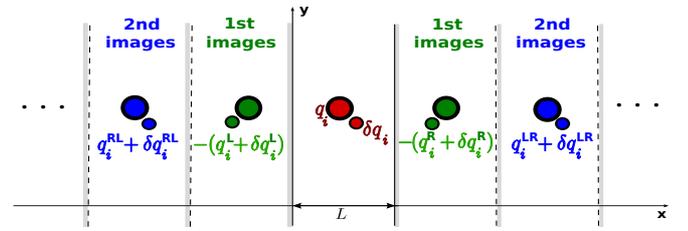}
\caption{Point charges between parallel plates, leading to an infinite series of image charges (first set of images in green, second set (images of images) in blue, \emph{etc}). Note the $\delta q_i$ added to the charges $q_i$ of the $N^\text{th}$ charge state, in going to the $(N+1)^\text{st}$ charge state: these also induce a series of repeated images.}\label{fg:imagemodel}
\end{figure}

Following Kaasbjerg and Flensberg,\cite{Kaasbjerg2008} and Mowbray \etal\cite{Mowbray2008}, we simplify the image-charge effects for the full spatial charge density by considering atomic point charges.
These are calculated from the charge states with $N-1$, $N$ (neutral) and $N+1$ electrons on the molecule. The atomic charges are denoted $q_j$, and are located at ${\bm r}_j$. The images of the atomic charges are denoted as $q^I_j$, and are located at ${\bm r}^I_j$; this position is found by (multiple) reflection with respect to the image planes.
\cite{Smith1989,Quek2007} When the total charge on the molecule changes, the atomic charges change by $\delta q_j$, inducing additional image charges $\delta q^I_j$ (see Fig.~\ref{fg:imagemodel}). The correction to a molecular level for a change in the charge state is then:
\begin{align}
\label{eq:JT}
\Delta = \sum_{i, j} \frac{\delta q_i q_j^I}{\rijI}
+ \sum_{i,j} \frac{\delta q_i \delta q_j^I}{\rijI} + U_\text{self}(\delta q_i)\;.
\end{align}

Eq.~\eqref{eq:JT} can be derived by considering the work needed to assemble the point-charge configuration.
The superscript $I$ implies a summation over the images. 
The first term is linear in the $\delta q_i$ and it represents the interaction between this added charge and the images $q^I_j$ of the reference configuration: this term affects the (constant) level shift.
The second term is quadratic in the $\delta q_j$ and it contains the interaction between the added charges and their images $\delta q^I_j$, and it is responsible for the gap renormalization. The last term collects the effects not depending on the molecule-metal separation. 

When there is only one image plane, and we neglect the self-interaction, the image-charge effect reduces to:
\[
\Delta_\text{ICE} = - \frac{2q\,\delta q + \delta q^2}{4z}\;,
\]
where we recognize the $1/(4z)$ potential shifts in the second term.

We study the image charge effects based on atomic charges calculated for the molecule \emph{inside the junction} (due to their nature as spatial decompositions, we prefer Hirshfeld or Voronoi decompositions over the basis-set decomposition involved in Mulliken decompositions) and compare the results with those based on the gas phase, as was done in the literature \cite{Mowbray2008,Hybertsen1986,Quek2007}.
In this way, we obtain the image-charge corrections to the occupied and unoccupied levels respectively for varying molecule-electrode distance.

To obtain the atomic charge distributions for different charge states of the molecule inside the junction, we use constrained density functional theory (CDFT). In CDFT, the minimum of the energy functional is searched under the constraint that the integral $\int f(\bfr) n(\bfr) \; d^3 r,$ has a pre-defined value. $f(\bfr)$ is a given function and $n(\bfr)$ is the electron density. We take $f(\bfr)$ to be 1 on the molecule, and 0 outside.

The constraint is ensured through a Lagrange parameter $V$ and is translated in to a term $Vf(\bfr)$ added to the potential. This extra potential, which is equivalent to a gate voltage (it is constant over the molecule), has been implemented in our transport code. 

In section~\ref{Sec:BDA}, we shall apply our method to a standard molecule and compare our results with those in the literature.
In section \ref{Sec:ZnTPP} we then apply our method for Zn-porphyrins, for which Perrin \etal performed  single-molecule experiments, and argue that the extra physics captured in 
our approach is essential for understanding the transport experiments.

\section{Image charge effects for benzenediamine (BDA)} \label{Sec:BDA}

\begin{figure}
\subfloat[BDA Gas-Phase Geometry]{\includegraphics[width=.475\columnwidth]{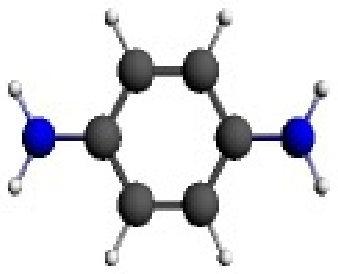}\label{BDA-gas-phase}}\hfill
\subfloat[Au-BDA Fragment Geometry]{\includegraphics[width=.525\columnwidth]{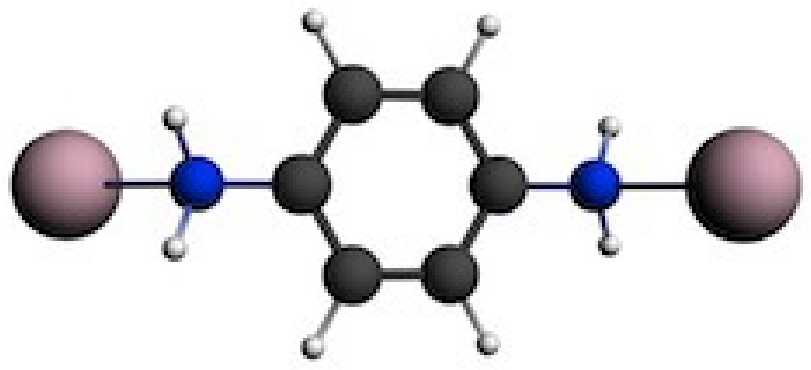}\label{BDA-fragment}}\\
\subfloat[Au-BDA Junction Geometry]{\includegraphics[width=.85\columnwidth]{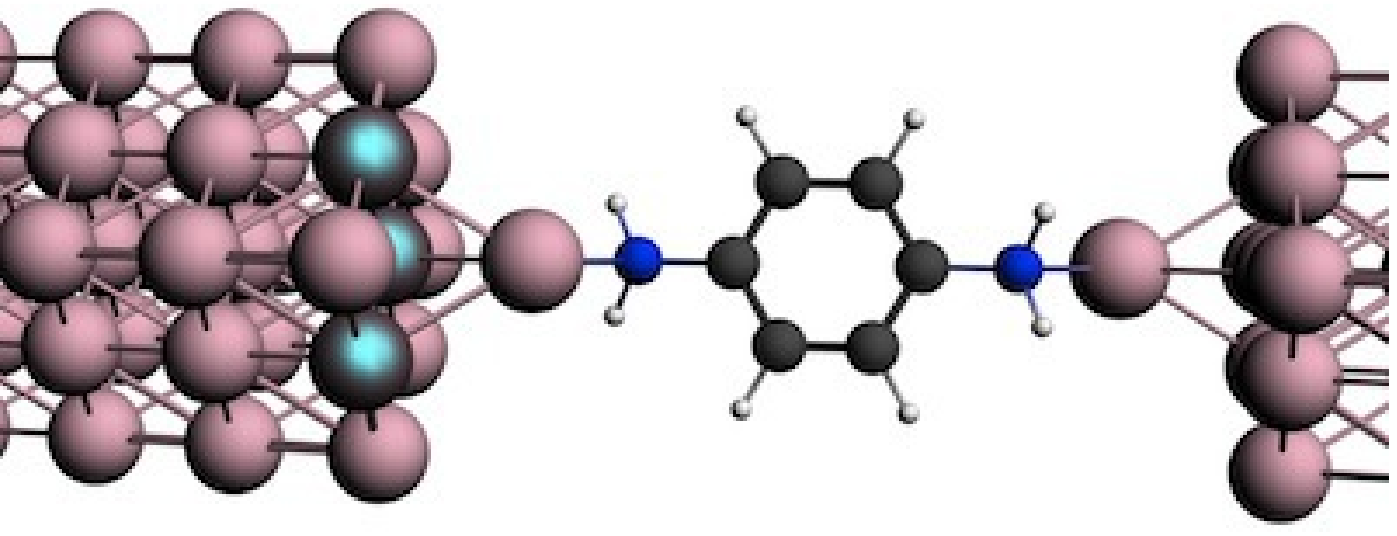} \label{BDA-binding}}
\caption{Geometries of BDA in (a) gas phase and (b) as a fragment. (c) (I,I) junction geometry. Metal ions are pink-grey, the blue-gray atoms are the substrate atoms coupled to the protruding gold atom. Left and right Au atoms show placement relative to a (111) surface.} \label{fg:3BDAgeometries}
\end{figure}

To analyze the transport through the molecule, we perform DFT-NEGF calculations for the Au-BDA-Au fragment attached to a FCC (111) surface (Fig.~\ref{fg:3BDAgeometries}). We consider a junction of type (I,I) according to the Quek \etal classification \cite{Quek2007}. For our calculations we use a TZP-basis of numerical atomic orbitals on the molecule, a DZ-basis of numerical atomic orbitals on the metal atoms and the GGA PBE functional in our implementation of NEGF-based transport in the ADF/Band quantum chemistry package \cite{Velde1991,Wiesenekker1991,Verzijl2012}. See supplemental material at [URL will be inserted by AIP] for further details concerning to the calculations. \footnote{See supplemental material at [URL will be inserted by AIP] for: Computational details in Section I; Explanation of where the gate field is applied in Section II; Comparison of the results obtained using LDA and GGA in Section III; Figure S1 shows the ZnTPPdT orbitals in gas phase; Figure S2 shows the regions where we applied the gate field used to determine the weakest coupling in the junction;  Figure S3 shows the levels shift obtained using LDA and GGA. Table I shows the Hirschfeld charge distribution for the BDA molecule.}

\begin{figure}
\subfloat[LUMO+1]{ \includegraphics[width=.5\columnwidth]{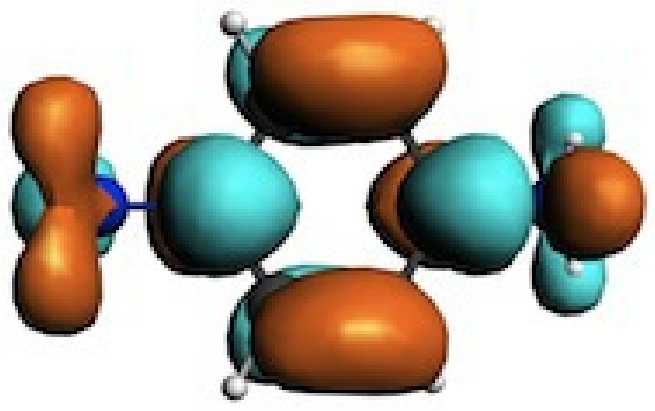} }
\subfloat[LUMO]{ \includegraphics[width=.5\columnwidth]{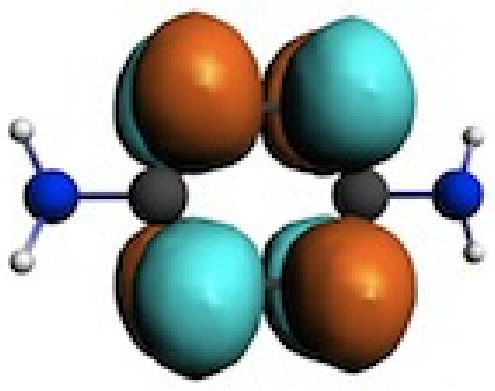} }\\
\subfloat[HOMO]{ \includegraphics[width=.5\columnwidth]{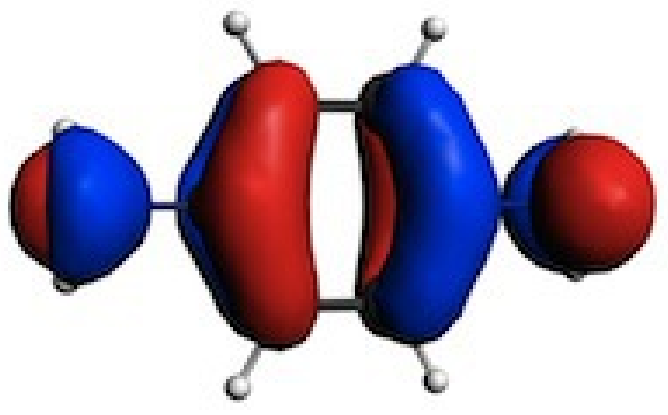} }
\subfloat[HOMO-1]{ \includegraphics[width=.5\columnwidth]{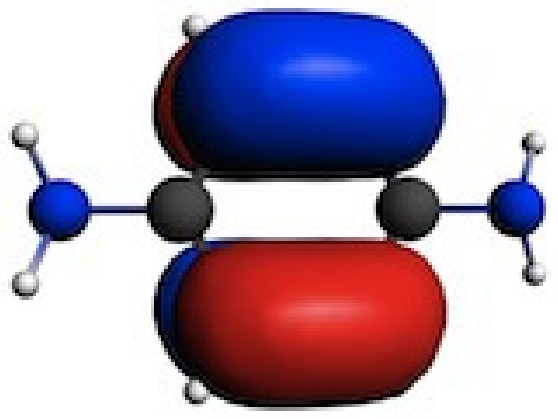} }
\caption{Orbitals of BDA molecule in gas-phase ordered by decreasing energy.}\label{fg:BDA-gas}
\end{figure}

\begin{figure}
\subfloat[LUMO+1]{ \includegraphics[width=.4\columnwidth]{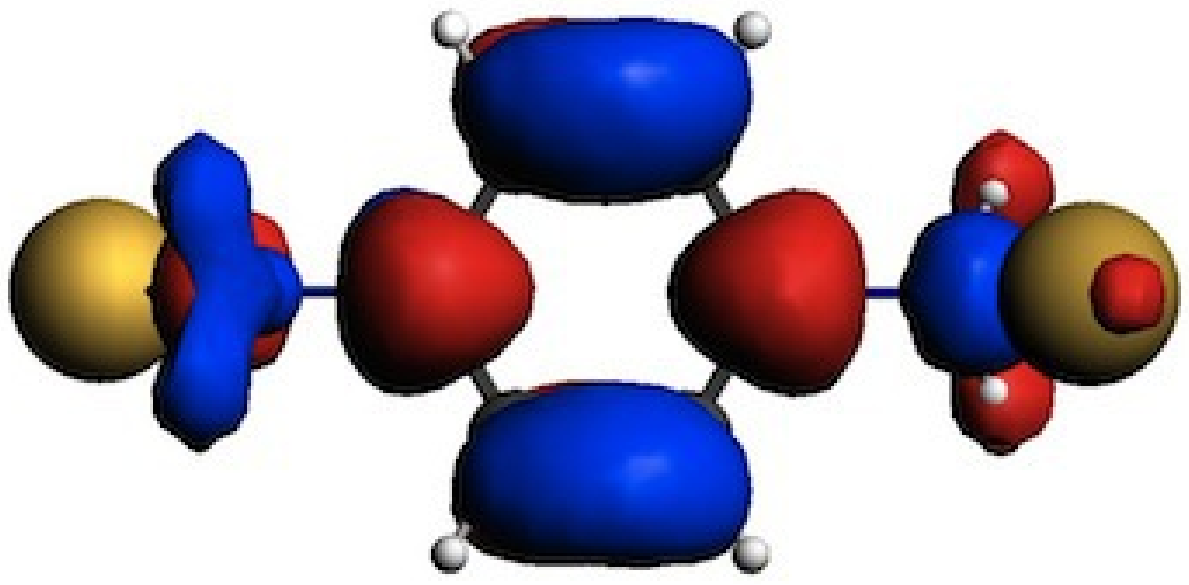} }
\subfloat[LUMO]{ \includegraphics[width=.4\columnwidth]{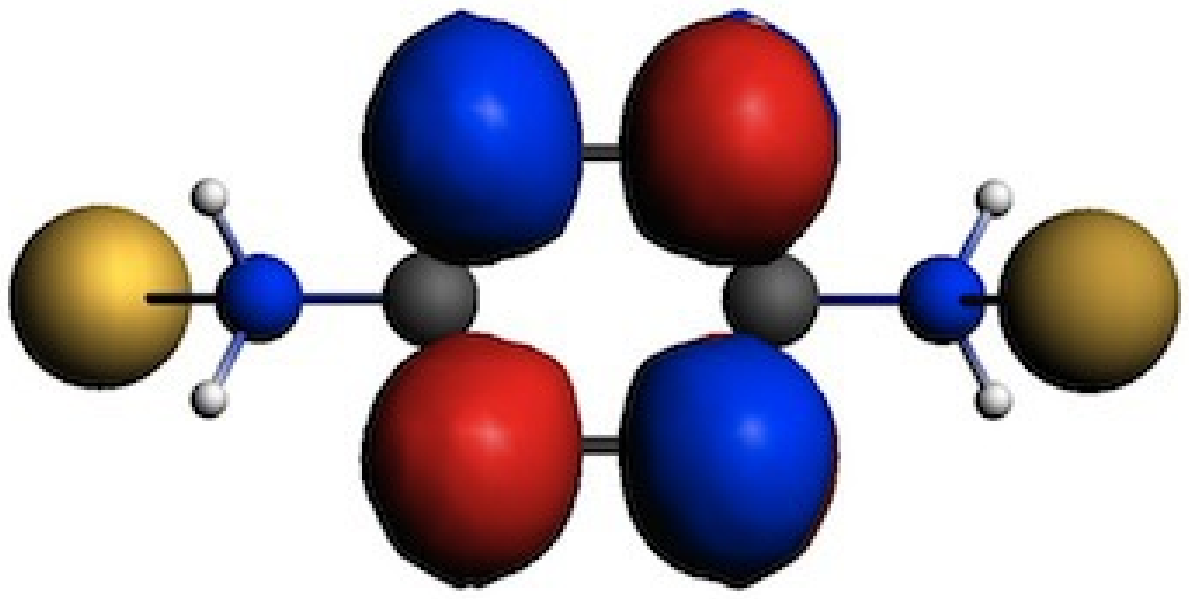} \label{fg:BDA-frag-LUMO}}\\
\subfloat[Interface state A]{ \includegraphics[width=.4\columnwidth]{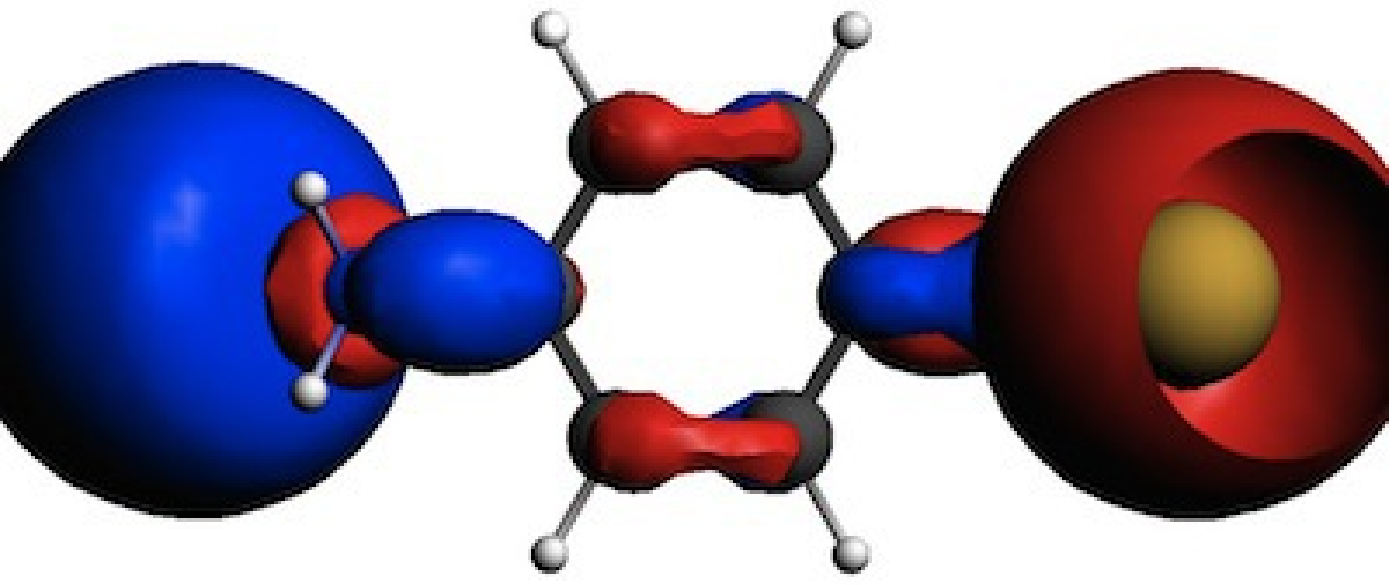}\label{fg:BDA-frag-gapA} }
\subfloat[Interface state B]{ \includegraphics[width=.4\columnwidth]{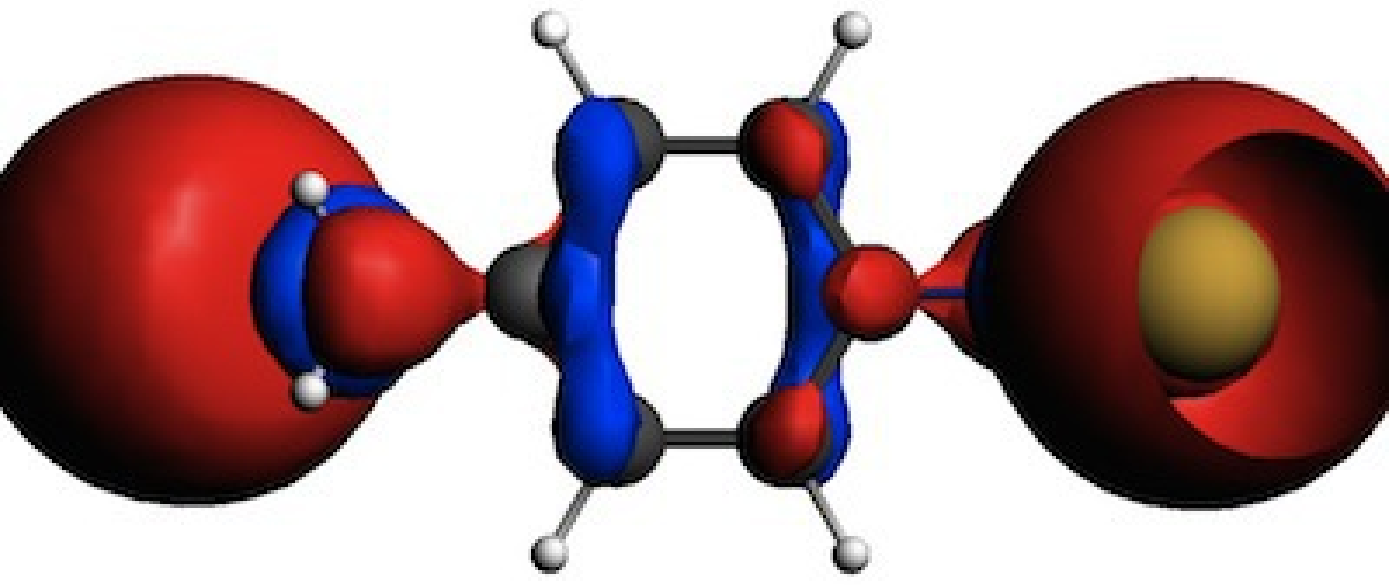}\label{fg:BDA-frag-gapB} }\\
\subfloat[HOMO]{ \includegraphics[width=.4\columnwidth]{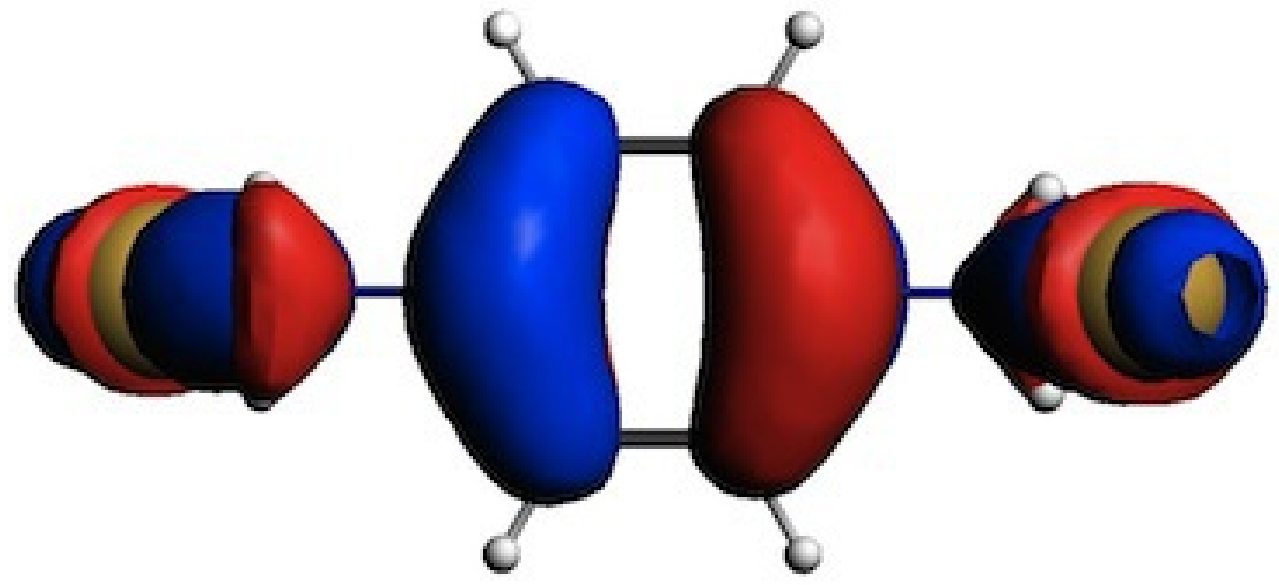} \label{fg:BDA-frag-HOMO}}
\subfloat[HOMO-1]{ \includegraphics[width=.4\columnwidth]{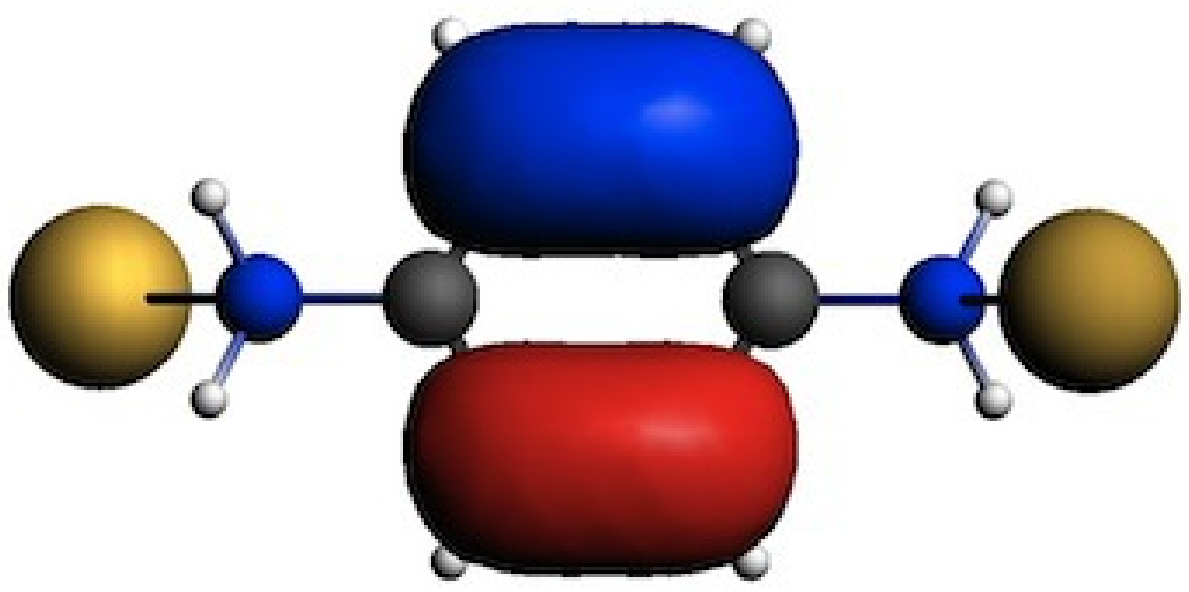} }
\caption{Au-BDA-Au Fragment orbitals labeled by their correspondence with the BDA gas phase orbitals (see Fig.~\ref{fg:BDA-gas}) and ordered by decreasing energy.}\label{fg:BDA-frag}
\end{figure}

For the molecule in the junction, we relax the geometry and we find the minimum energy configuration. Then, we stretch the junction separating the contacts with the molecule's conformation unchanged.

\begin{figure}
\subfloat[Molecule close to the contacs (stronger coupling).]{ \includegraphics[width=.65\columnwidth]{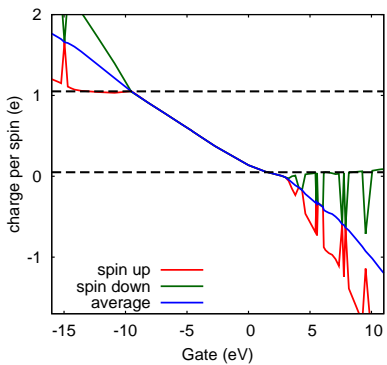} \label{fg:Molecule-close}}\\
\subfloat[Molecule far from the contacts (weaker coupling).]{ \includegraphics[width=.65\columnwidth]{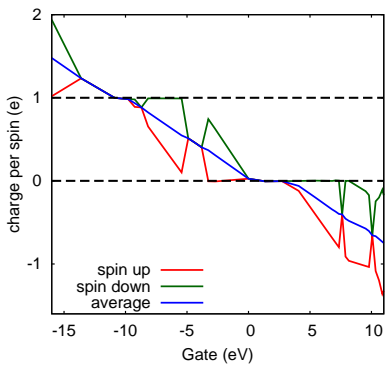} }
\caption{Spin-resolved occupation as a function of the applied gate when (a) the molecule is close to the contacts and (b) the molecule is far from the contacts.}\label{fg:spin-resolv}
\end{figure}

The spin-resolved occupation (see Fig. \ref{fg:spin-resolv}) indicates how the filling of the individual levels changes upon varying the gate. We have calculated the spin-resolved occupation for two cases: one where the molecule is close to the contacts (we consider the energy minimum for this, see Fig. \ref{BDA-binding}) and one where the molecule is far away. 
Sometimes, we see spin-polarization, unless the occupation happens to be an even integer. We also observe absence of this polarization in the strong coupling limit for the charge between 0 and 1. 

We expect the presence of polarization to be related to the weak coupling condition $\Gamma < U$, where $U$ is the Coulomb repulsion for electrons at the relevant level. Polarization is not expected when $\Gamma > U$. This appears to be the case in the short distance configuration (strong coupling limit) when the charge is between 0 and 1.

We emphasize that this polarization is not physically correct as the system has unpolarized leads -- hence the chemical potentials for both spin directions are identical. However, it has been pointed out by several researchers that the spin-polarized states found in DFT calculations can give us valuable information about the levels and their occupation \cite{FLiu2013,Bergfield2012,ZFLiu2015,Burke2012}.

For the weak-coupling case, we see `plateaus' occurring in the levels corresponding to fixed occupation demonstrating that only one type of spin is added to or removed from the system when changing the gate. These plateaus are sometimes interrupted by unpolarized points; we assign these to anomalies in
the self-consistency cycle.

In the stronger-coupling case, we also see plateaus, although they are less flat, and, importantly, they do not correspond to integer occupation, but slightly above
. Apparently there is some `extra' charge on the molecule in these cases -- however, the deviations may also be related to the (Hirschfeld) calculation of the atomic charges. We conclude from figure \ref{fg:Molecule-close} that there is a constant background charge on the molecule, corresponding to $+0.05e$ per spin (see dashed line). The fact that the two easily identifiable plateaus (red curve around $-10~eV$ and green curve around $+5~eV$) are separated by (very nearly) 1e per spin, indicates how charges should be added or removed from the reference state.

\begin{figure}
\includegraphics[width=.8\columnwidth]{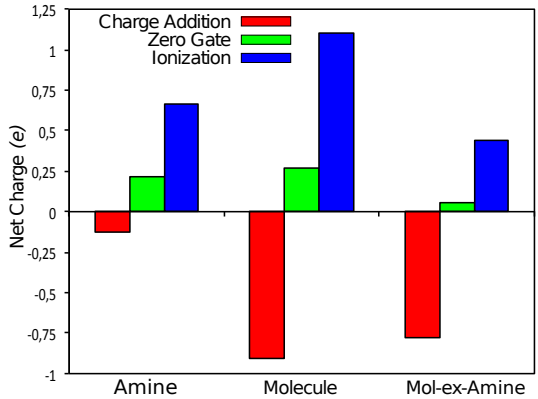}
\caption{Hirshfeld projected charges for the three gated transport levels (the reference state and $\approx\pm e$ charged states), showing the difference in charging the molecule, amine groups and molecule-without-amine as the gate field is varied.} \label{fg:gated_charges_BDA}
\end{figure}

The reference charge of the molecule in the junction, for the relaxed geometry, is $+0.274e$. This is due to \emph{both} spin directions -- therefore we have a charge of $+0.137e$ per spin. In this state, the molecule has already some extra charge due to partial charge transfer across the interface \cite{Thygesen2009}. This is a charge excess of $+0.087e$ with respect to the background charge $+0.05e$. In order to remove one electron from the molecule, we therefore need to add $+0.913e$ and to put an extra electron corresponds to $-1.087e$ (see Fig.~\ref{fg:gated_charges_BDA}). 

\begin{figure}
   \includegraphics[width=\columnwidth]{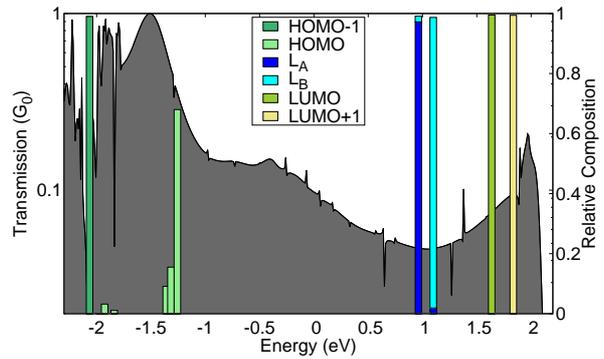}
\caption{Peaks Decomposition with fragment Orbital Levels (grey shaded curve is transmission). Composition of peaks in transport, constructed by projection onto fragment molecular orbitals. A state with value 1 is a decoupled state, and completely un-hybridized (\eg HOMO-1), while HOMO, is strongly hybridized, with the rest originating from Au).}\label{fg:BDA-peak-composition}
\end{figure}

Fig.~\ref{fg:BDA-peak-composition} shows the compositions of the peaks in the transmission through the Au-BDA-Au junction near $\epsilon_{f}$. For these, we project the eigenstates of the transport calculation onto the orbitals of the Au-BDA-Au fragment \cite{Verzijl2012}.

The HOMO projection is composed of many such orbitals as a result of the hybridization with Au, in contrast to the LUMO, the LUMO$+1$ and the HOMO$-1$ states. The HOMO and LUMO$+1$ are more dominant in charge transport than the LUMO and HOMO$-1$ states which contribute weakly due to their strong localization at the center of the molecule. The interface states ${A}$ and ${B}$ do not show up as peaks in the transmission due their very low density at the center of the molecule -- see Fig. \ref{fg:BDA-frag-gapA} and Fig \ref{fg:BDA-frag-gapB}. 

\begin{figure}
\subfloat[Geometry for Image-Charge Shifts]
{
   {\includegraphics[width=.5\columnwidth]{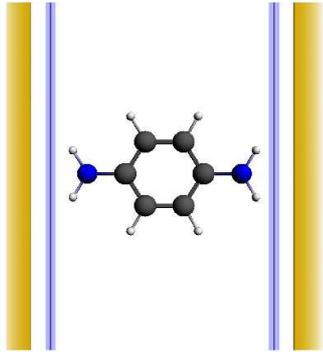}\label{fg:shiftsgeomBDA}}
}\\
\subfloat[Transport Gap Renormalization]{
   {\includegraphics[width=0.9\columnwidth]{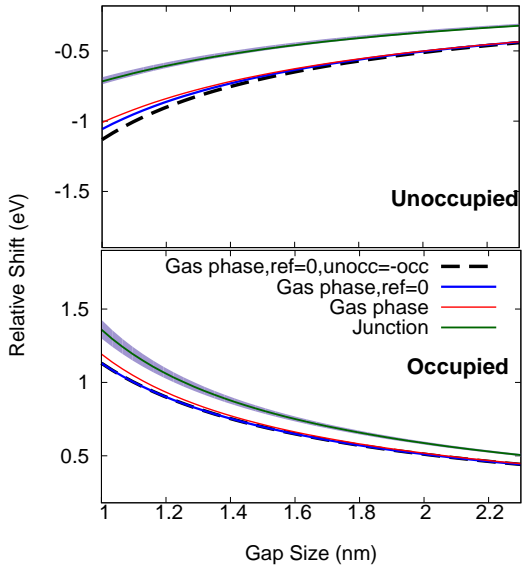}\label{fg:shiftscalcBDA}}
}
\caption{(a) Geometry used in the image-charge model (with uncertainties) (b) Comparison of results for total image-charge corrections using charges from gas phase calculations of BDA and from the molecular junction as a function of the distance between the contacts.
}\label{fg:contactsBDA}
\end{figure}

We now consider the results for the calculation of the image charge effect. 
In Fig.~\ref{fg:shiftsgeomBDA} we show the geometry used for our image-charge calculation and in Fig.~\ref{fg:shiftscalcBDA} we show the resulting shifts of the occupied and unoccupied levels as function of the distance between the two contacts. The uncertainty bands are calculated based on a $\pm0.25\ \AA$ uncertainty in the position of the image planes.
In Fig.~\ref{fg:shiftscalcBDA}, we also compare the results obtained by our method using different assumptions during the calculations. The dashed line is calculated using the gas phase charge distribution, zero charge on each atom for the reference state and omit the atomic charges associated with the EA, following the Mowbray \etal assumptions. 
Using \emph{different} charges for the calculation of the image charge effect of the occupied level (blue line), the symmetry for the shifts of the occupied and unoccupied levels is not maintained, although the curves remain close. Using the charge distribution of the neutral molecule as reference state (red line), these differences increase slightly. Finally, using the junction charge distribution (green line), results in a substantial difference. This shows that by using charges obtained with the junction geometry (from a NEGF+DFT calculation) for the image-charge calculation, we are including features that are absent when the gas phase charges are used.

\section{Au-ZnTPPdT Molecular Devices} \label{Sec:ZnTPP}

We now proceed to a more complicated application of the method, which allows for comparison with a recent experiment that revealed the image-charge effects on both occupied and unoccupied molecular levels.

\subsection{Experimental Results}\label{experiments}

We consider the experimental findings of Perrin \etal\cite{Perrin2011a,Perrin2011b,Perrin2013} who studied thiol-terminated zinc-porphyrin molecules [Zn(5,15-di(p-thiolphenyl)-10,20-di(p-tolyl)porphyrin)], abbreviated as ZnTPPdT. 
In the experiments, the current was recorded as a function of gate and bias voltage, and of the electrode separation.
Peaks in the differential conductance were identified as transport resonances.
These resonances show a marked ``mechanical gating'' effect, where a level shift is induced by a change in the metal-molecule distance (for both the occupied and unoccupied levels of the molecule). The efficiency of the effect can be expressed by a mechanical gate coupling (MGC) defined as 
\begin{equation}
\epsilon_F=\frac{d V_b}{dx}, 
\end{equation}
where $V_b$ is the bias voltage and $x$ the electrode separation . 

\begin{figure}
\includegraphics[width=.8\columnwidth]{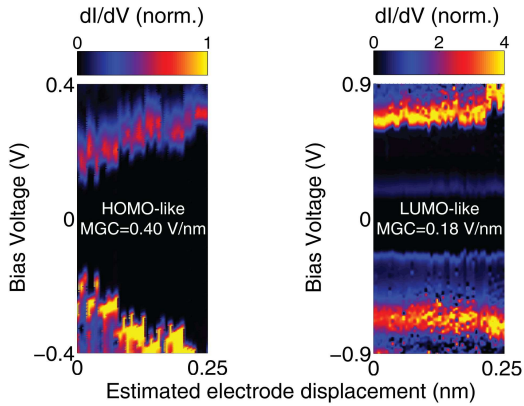}\label{fg:exp_measurement}

\caption{Representative measurement,\cite{Perrin2013} showing HOMO-like and LUMO-like observed MGC's. Note the dilation of the y-axis in the case of LUMO-like resonances. 
}
\label{fg:experiment}
\end{figure}

We show experimental data for these shifts in Fig.~\ref{fg:experiment}, where the measurements show a distance-dependent energy for the lowest resonance. A linear fit of the resonance positions was used to find the MGC. 
The broadness of the distribution is presumably due to the fact that ZnTPPdT is not a rod-like molecule; it can form molecular junctions with various geometries, as has been reported previously for similar molecules.\cite{Perrin2011b}

The MGC's values are 
in the range of $0.2-1$ $V_b$/nm 
Combined with a typical range of $0.5$ nm over which the junctions formed are stable, 
implying levels shift of roughly $50-250$ meV in energy, if we assume the bias voltage  drops symmetrically. An average MGC values of $0.40$ V/nm was found for occupied, and  $0.18$ V/nm for unoccupied levels.

\subsection{Calculations}\label{calculations}

We will now show that our approach yields trends matching the experiment, and explains the asymmetry in the shifts found between occupied and unoccupied levels. 

\begin{figure}
\subfloat[Interface State A]{ \includegraphics[width=.4\columnwidth]{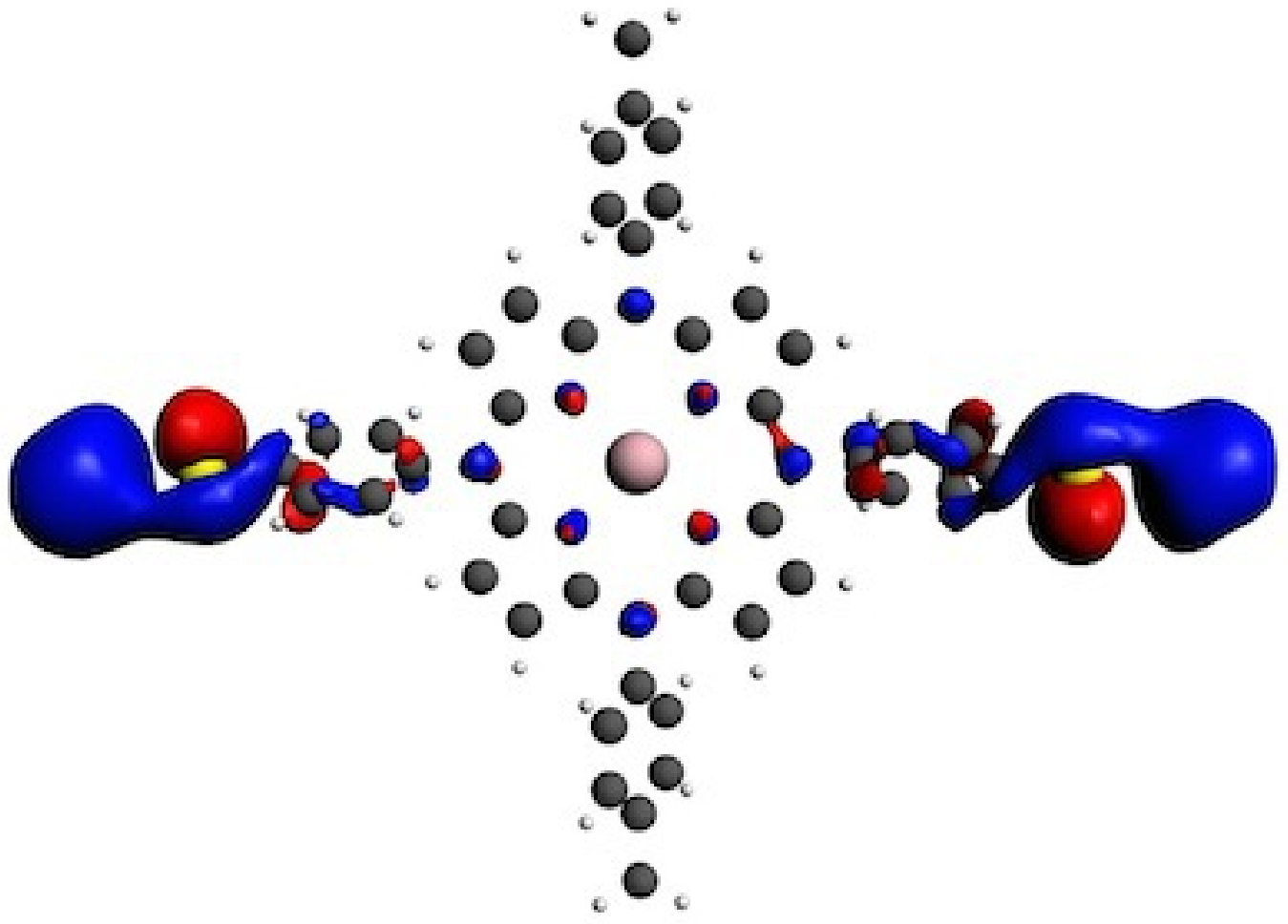} }
\subfloat[Interface State B]{ \includegraphics[width=.4\columnwidth]{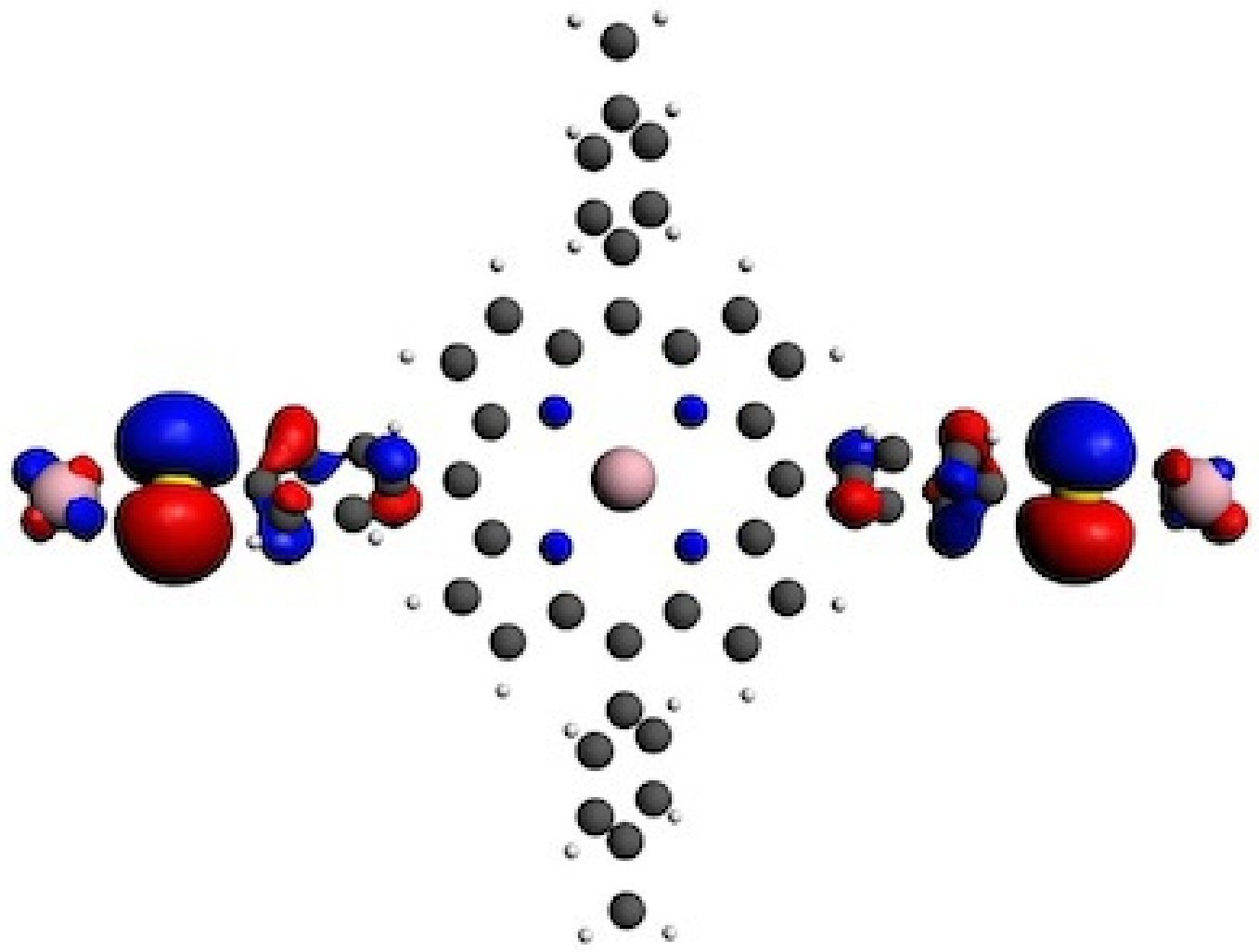} }\\
\caption{Typical interface levels which form on hybridizing with Au: 6 total between the analogue of the gas phase HOMO and LUMO.}\label{fg:ZnTPP-frag}
\end{figure}

We focus on the frontier orbitals (HOMO and LUMO) which are generally considered to be the most useful for transport. We find a the HOMO-LUMO gap to be $1.8$ eV in our LDA and GGA calculations and $2.7$ eV using the B3LYP functional, consistent with the reports of Park \etal\cite{Park2008}, and in general agreement with their redox measurements of roughly $2.2$ eV.  

Our Au-ZnTPPdT binding geometry is based on a phenyl ring bonded to an FCC (111) gold surface via a thiolate bond, in a hollow-site configuration.\cite{Nara2004,Andrews2006,Kondo2006,Pontes2011}

In the calculations, the binding is characterized by chemisorption, with significant charge transfer to the thiols, which act as acceptors. This is in agreement with the literature on such bindings. \cite{Xue2003a,Xue2003b,Love2005,Hoft2006,Romaner2006} 

All calculations were performed using a TZP-basis of numerical atomic orbitals on the molecule, using the LDA functional  with thiols located at a 2.59\AA\xspace from the electrodes. 

In figure~\ref{fg:ZnTPP-frag}, we show two interface orbitals of the ZnTPP-fragment, which contains two extra gold atoms. There are six such states, in addition to the direct counterparts of the LUMO, HOMO and HOMO-1 of the gas phase ("See supplemental material at [URL will be inserted by AIP] for the ZnTPPdT orbitals in gas phase." ). Of these six, two pairs relate to the HOMO, and one to the LUMO. The orbital levels in these fragment pairs appear to be of a bonding/anti-bonding character, with splittings on the order of $0.1$~eV.

\begin{figure}
\subfloat[Peaks Decomposition with Molecular Orbital Levels (grey shaded curve is transmission)]{
   \includegraphics[width=0.95\columnwidth]{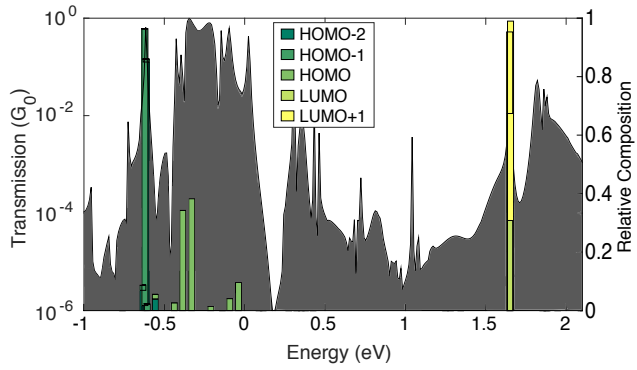}\label{fg:decompositions}
   }\\
\subfloat[Peaks Decomposition with Interface Levels]{
   \includegraphics[width=0.95\columnwidth]{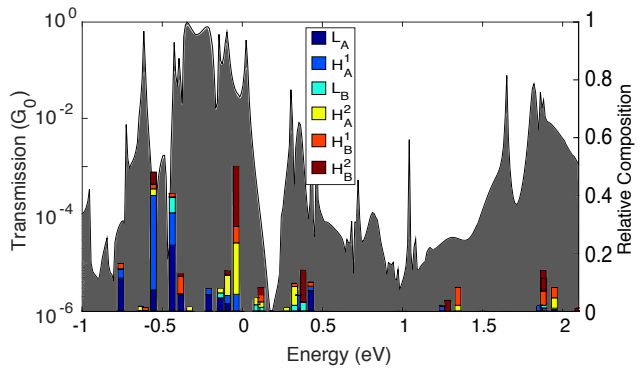}\label{fg:gaplevels}
   }
\caption{(a) Composition of peaks in transport, constructed by projection onto fragment molecular orbitals. A state with value 1 is a decoupled state, and completely un-hybridized (\eg HOMO-4 through --7), while HOMO-1, --2 and HOMO are strongly hybridized with each other \emph{and} the Au electrodes (reflected in their 30-50\% representation in the junction levels, with the rest originating from Au). The LUMO and LUMO+1 peaks are likewise strongly mixed with each other, coupling much less to the Au, reflected in the much narrower transport peaks near $1.7$ eV. (a) As in (a) for the interface levels rather than for the molecular orbitals shown in (a). }\label{peak-compositions}
\end{figure}

\begin{figure}
\includegraphics[width=.8\columnwidth]{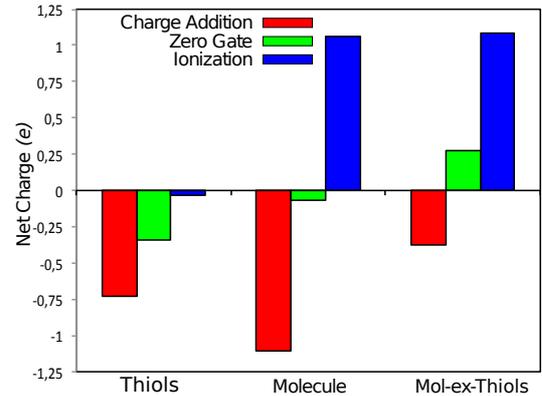}
\caption{Partial charges for the three gated transport levels (the reference state and gated such that the net charge is $\approx\pm e$), showing the difference in charging the molecule, thiols and molecule-without-thiols as the gating is varied. At zero-field, the molecule is roughly neutral, with negative thiols and a positive core.} \label{fg:gated_charges}
\end{figure}

Fig.~\ref{peak-compositions} shows the transmission of a typical transport calculation for the MCBJ geometry.

We observe a cluster of HOMO-like peaks near $\epsilon_f$ (defined as $0$ eV), some small peaks inside the gap near $0.4$ eV, and the nearly-degenerate LUMO and LUMO+1 around $1.7$ eV. Fig.~\ref{fg:decompositions}, shows the decomposition of the transmission into fragment orbitals directly corresponding to molecular orbitals, and in Fig.~\ref{fg:gaplevels} for the interface orbitals.

The peaks right below the Fermi level derive mostly from the HOMO,\footnote{Identified by analyzing the orbital symmetries of the wavefunctions of these levels.} with significant amounts of interface levels mixed in.
Fig.~\ref{fg:gaplevels} shows the role of the 6 interface levels labeled L$_\text{A,B}$, H$^1_\text{A,B}$ and H$^2_\text{A,B}$, derived from hybridization of HOMO and LUMO with the gold. 

For ZnPPTdT, the level splitting between the interface states is extremely small. This means that there is not a unique state going to be filled, and this precludes spin polarization and plateaus like those in Fig. \ref{fg:spin-resolv} are absent. On the other hand, the total charge in the reference state is only $0.05e$, distributed over the two spin directions, and this will therefore contribute only very slightly to the difference between the curves for occupied and unoccupied levels. We therefore just add or subtract one unit charge in order to find the reduced and oxidized states. 

\begin{figure}
 \subfloat[]{\includegraphics[width=.6\columnwidth]{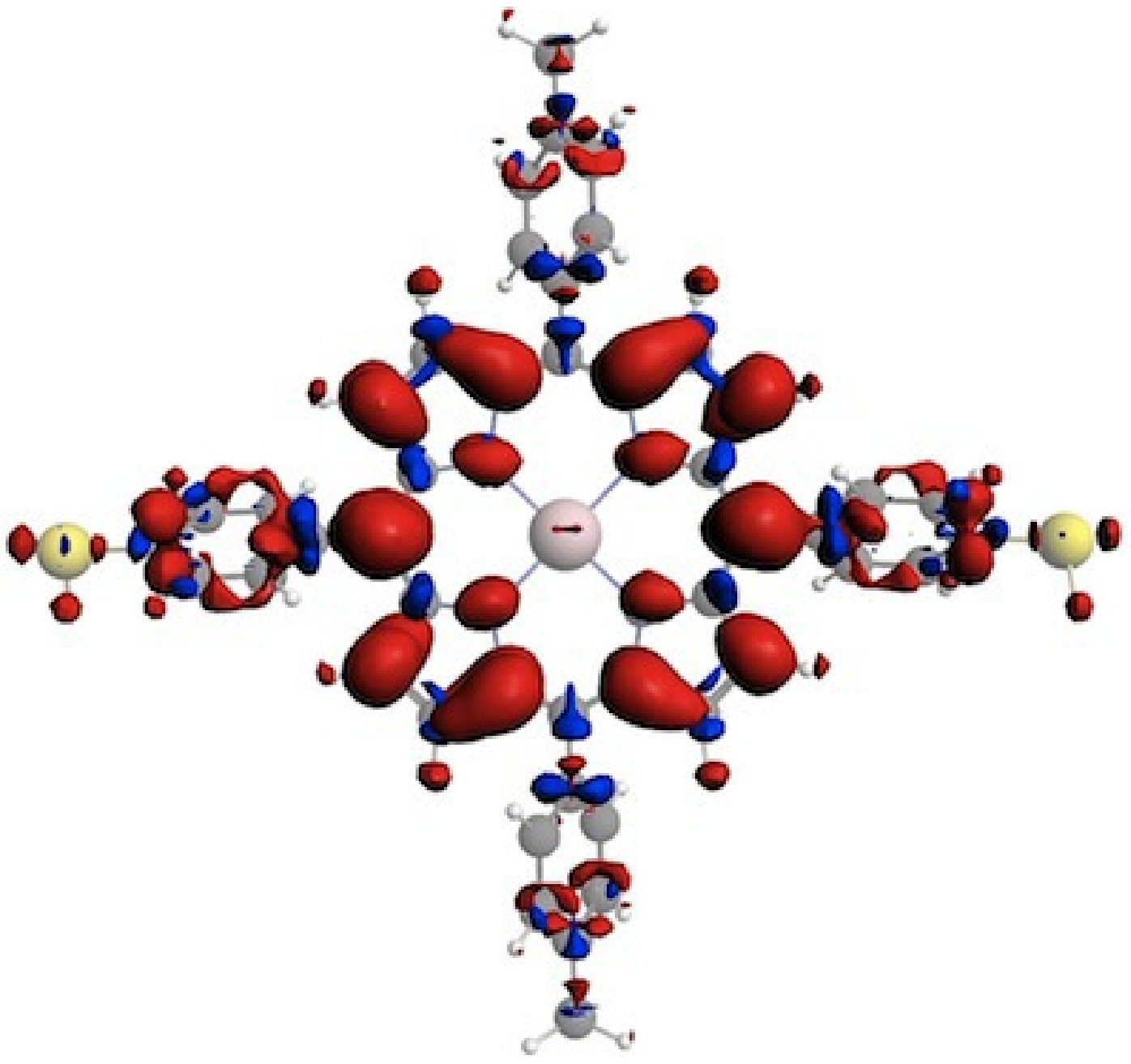}}\\
 \subfloat[]{\includegraphics[width=\columnwidth]{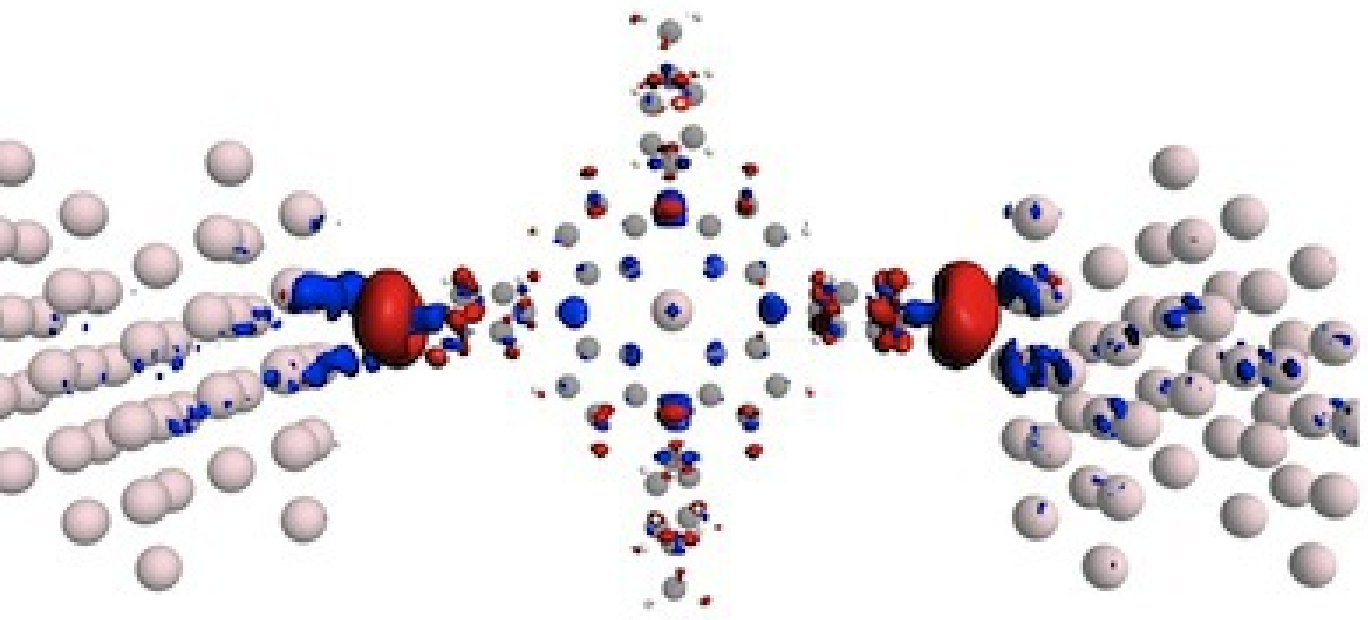}}
\caption{Difference in charge distribution in the $N+1$ relative to the $N$ electron charge states. Red indicates the increase of negative charge when adding an electron; blue the decrease. Differences for (a) gas phase DFT calculations (LUMO like difference) and (b) for gated DFT+NEGF transport calculations (recalling the interface levels of Fig.~\ref{fg:ZnTPP-frag}).} \label{fg:NNpm1-charges}
\end{figure}

We have applied our method for calculating image-charge effects to this junction. In the reference state the net charge is $-0.05e$ with a strongly negative charge ($-0.34e$) on the thiols, and $+0.29e$ on the rest of the molecule, mainly on the Zn ion. Figure~\ref{fg:NNpm1-charges} shows the 
difference in charge for the ionized ($N-1$) states with respect to the reference state. This difference 
resides mostly on the arms, increasing the image charge effect a lot due to the proximity of the extra charge to the contacts. 

The fact that in the reference state, the charge on the thiols is approximately the opposite of the 
charge on the rest of the molecule, is responsible for a significant difference in slope for the occupied and
unoccupied levels.


\begin{figure*}
\subfloat[Geometry for Image-Charge Shifts]{
   \raisebox{0.8cm}{\includegraphics[width=.9\columnwidth]{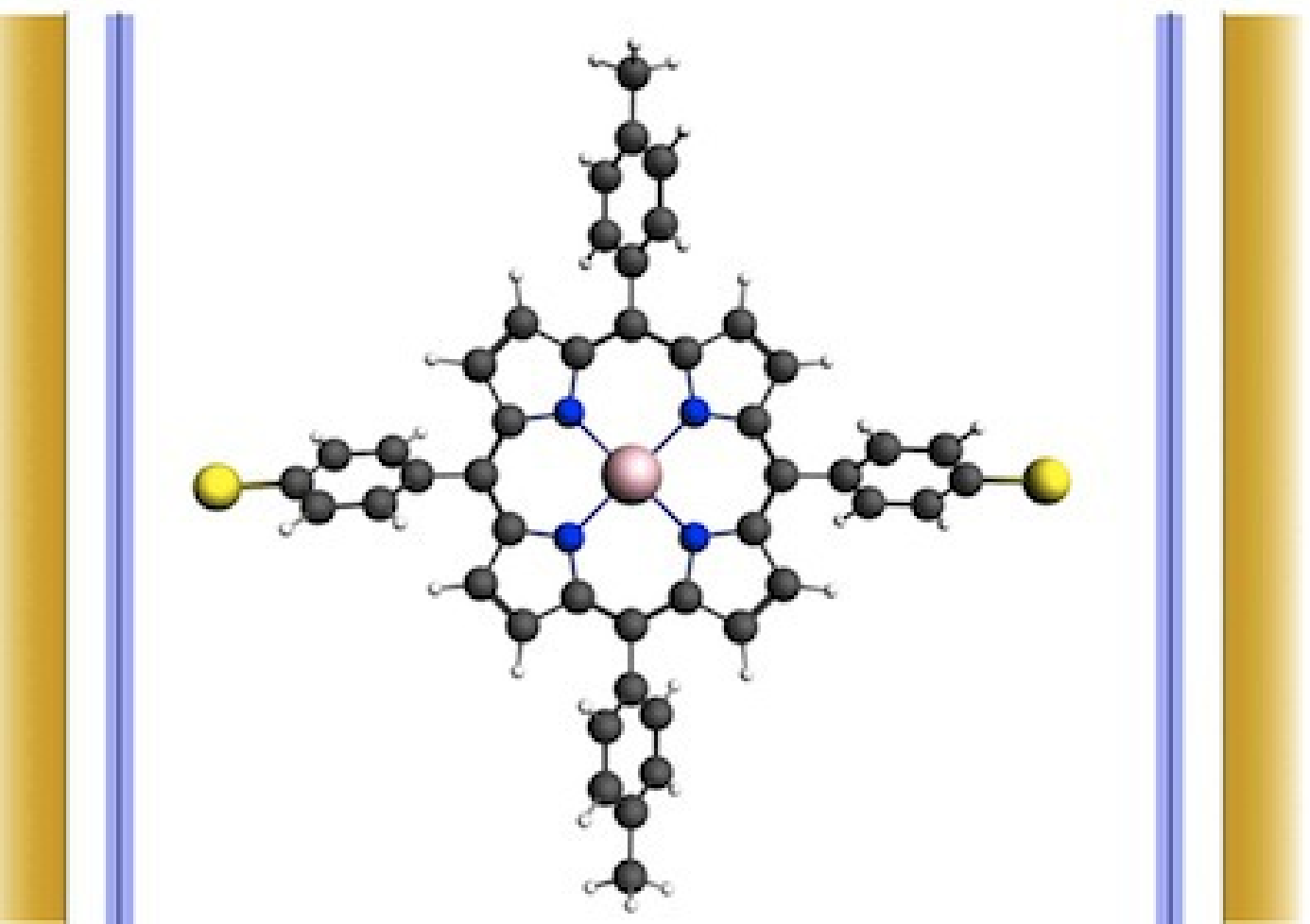}\label{fg:shiftsgeom}}
   }
\subfloat[Transport Gap Renormalization]{
   \includegraphics[width=.8\columnwidth]{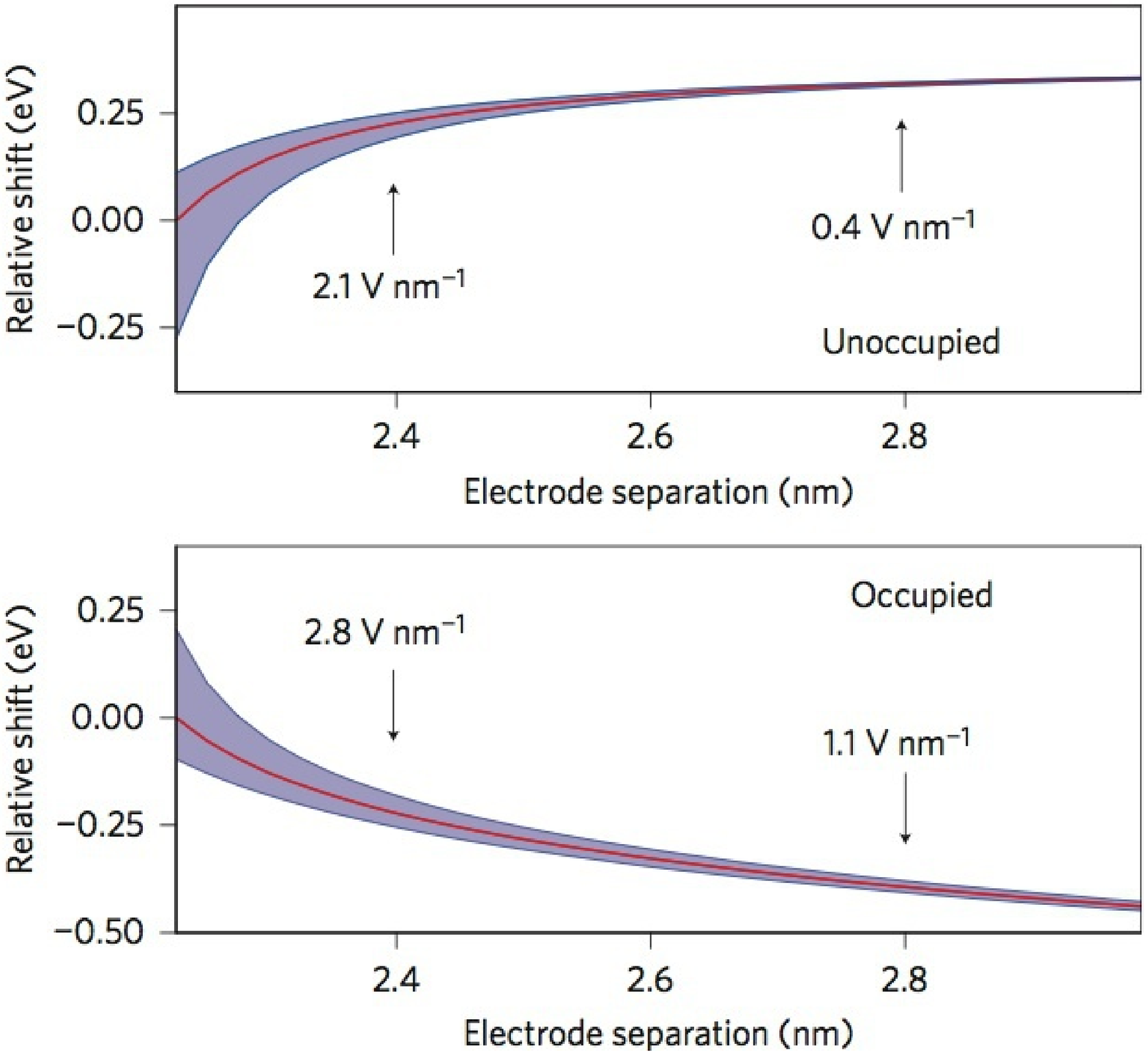}\label{fg:shiftscalc}
   }
\caption{
(a) Geometry used in the image-charge model, and (b) shifts predicted by the model (with uncertainties) showing the occupied- and unoccupied-levels both shifting towards $\epsilon_F$ with MGC's (the derivative with distance) in the range of $0.2-1.4$ eV/nm, expressed in the symmetrically applied bias. 
}\label{fg:Shifts}
\end{figure*}

The calculated level shifts as a function of distance are plotted in Fig.~\ref{fg:shiftscalc} 
Our calculations predict MGC's in the range of 
$1.1-2.8$ eV/nm for an occupied level and
$0.4-2.1$ eV/nm for an unoccupied level 
(in opposite directions), depending on the electrode separation (see Fig.~\ref{fg:shiftscalc}).
The different slopes differ significantly indeed, confirming the experimental findings. 

To obtain this difference, a detailed calculation of the molecule inside the junction is essential. Using gas-phase orbitals, the wrong orbital
(LUMO) would have been chosen as the unoccupied transport level, and the substantial contribution of the charge located at the arms of the hybridized 
HOMO would have been missed. 

Our calculations reveal that the background image-charge effect contributes significantly to the MGC and explains the distance-dependent renormalization of the position of the molecular orbital levels with respect to the Fermi level of the electrodes. 
Taking the reference state to be the gas phase neutral state suppresses the asymmetry between the shifts for occupied and unoccupied levels. 
This supports our conclusion that for the measurements of Fig.~\ref{fg:experiment} an interface-stabilized level of the fragment has lost some charge, as is suggested by the peak above the Fermi level in our transport calculations, and that this level is being addressed in electron transport through the unoccupied state.

\section{Conclusions}

In summary, we have presented a method for calculating the image-charge effects which change the alignment of the occupied and unoccupied levels in molecular devices with the Fermi levels of the electrodes. Our approach is based on the charge distribution of the molecule in the junction in different charge states. It is essential to use these rather than their gas phase equivalents for two reasons. First, the relevant charge states may have a different character in gas phase molecules and molecules in a junction, due to the formation of ``interface levels'' in the latter. These are stabilized by the metal-molecule interface, and have no counterpart in the gas phase. Second, unlike in the gas phase, the reference state in the junction (at zero bias and gate) can carry a net charge, which implies a significant contribution to the reduction of the metal work function upon chemisorption of a molecule.

We have applied our method to a standard benzenediamine molecule and found results in good agreement with those obtained using Mowbray's \etal model. The results differ however in our approach, mainly
due to the nonzero charge in the reference state, and because we also address interface states that differ essentially from gas level orbitals.

Perrin \etal's\cite{Perrin2013} experiments on Au-ZnTPPdT reveal distance-dependent level shifts which are in agreement with our calculations. In this experiment, the fact that the reference state is non-neutral causes the MGC for occupied and unoccupied levels to be quite different. 
Our model agrees with the experimentally determined shifts within a factor of two.

Our approach demonstrates that for addressing image-charge effects within DFT, considering molecules in the junction is essential.

\begin{acknowledgements}
The authors thank the financial support by the Dutch Foundation for Fundamental Research on Matter (FOM), the EU FP7 program under the ``ELFOS'' grant agreement and a grant by the Netherlands' National Computing Facilities Foundation, financed by The Netherlands Organization for Scientific Research (NWO). We also thank C.A. Martin, J.S. Seldenthuis, F. Grozema, R. Eelkema and J. van Ruitenbeek for fruitful discussions.
\end{acknowledgements}

\bibliography{Image-Effects-in-Transport-Calculations}

\end{document}